\documentclass[pdftex,twocolumn,epjc3]{svjour3}          

\RequirePackage[T1]{fontenc}

\smartqed  

\RequirePackage{graphicx}

\RequirePackage[numbers,sort&compress]{natbib}
\RequirePackage[colorlinks,citecolor=blue,urlcolor=blue,linkcolor=blue]{hyperref}
\RequirePackage{amsmath,graphicx,csvsimple,booktabs,tikz,amssymb,physics,bm, slashed}

\journalname{Eur. Phys. J. C}

\newcommand{\der}[2]{\frac{d#1}{d#2}}
\newcommand{\partialder}[2]{\frac{\partial #1}{\partial #2}}
\newcommand{\partialfunc}[2]{\frac{\delta #1}{\delta #2}}
\newcommand{\Hamiltonian}[0]{H[\phi, \pi]}
\newcommand{\Wigner}[0]{W[\phi, \pi]}

\begin{document}

\title{Functional Quantum Field Theory in Phase Space}

\author{José A. R. Cembranos\thanksref{e1,addr1}
        \and
         Marcos Skowronek\thanksref{e2,addr1} 
}

\thankstext{e1}{e-mail: cembra@ucm.es}
\thankstext{e2}{e-mail: marcosko@ucm.es (corresponding author)}

\institute{Departamento de Física Teórica, Universidad Comlputense de Madrid, 28040 Madrid, Spain\label{addr1}
}

\maketitle

\tableofcontents

\vspace{0.5 cm}

\begin{abstract}
We determine the form of the Wigner functional for several types of quantum free field theories in order to analyze the representation of QFT in phase space, as well as to compare it to other mainstream formulations. We use Jackiw's functional representation of a quantum field state and extrapolate it to the phase space formalism using an extension of Moyal's equation.
\end{abstract}

\section{Introduction}

The laws of Classical Mechanics are well established since the seventeenth century. However, although Newton's laws describe the world correctly within their range of application, they are ineffective when dealing with more complex problems. In the nineteenth century, other formulations emerged, such as those of Lagrange and Hamilton, which constitute a much more elegant description, treating positions and momenta equally, i.e. within the phase space.

Quantum Physics emerged at the beginning of the 20th century trying to describe specific phenomena. Its formulation began to be developed in the 1920s as a description of the subatomic world. In 1926, Schrödinger formulated wave mechanics, where the central concept was the wave function. This wave function  represents an element of an infinite-dimensional Hilbert space, which contains the possible states of the system. Its squared modulus represents the probability of finding a particle at a given point. In 1927, Heisenberg enunciated the Principle of Uncertainty, which places a limit on the accuracy of certain observations. Specifically, it ensures that it is not possible to measure the momentum and position of a particle with absolute precision. This seems to prevent the formulation of Quantum Physics within phase space, since the operators associated with both observables do not commute. However, this is not the case.

The first attempt of defining quantum physics in phase space was made by Weyl in 1927 \cite{Weyl}. In this pioneer work, he establishes a relationship between functions defined in phase space and {\it Weyl} ordered operators. Such order is defined as completely symmetric for non-commuting position and momenta operators defined in Hilbert space. He thought that it was the bridge that linked the classical and the quantum, although later it was shown that it was just a change of representation. In 1932, Von Neumann postulated the correspondence rules between Quantum and Classical Physics \cite{neu}. Also in 1932, Wigner introduced his homonym function in a different context, attempting to calculate quantum corrections to thermodynamic equilibrium \cite{Wigner_function}. This function can be understood as a generalization of the Liouville density, but it cannot be interpreted as a classical probability distribution, since it can take negative values. Based on it, in the 1940s, Groenewold and Moyal developed a formulation of Quantum Mechanics in phase space, where a non-commutative product appears, the star product, and the evolution of the Wigner function occurs through from a deformation of the Poisson bracket: the Moyal bracket \cite{Groenewold,Moyal}.

At present, the use of the Quantum Mechanics formulation in Phase Space is restricted to very specific areas. On the other hand, its generalization to Quantum Field Theory (QFT) is not straightforward. It requires to make decisions about different assumptions and subtleties. Throughout this work, we will develop the functional formalism of QFT, where the states are represented by functionals of fundamental fields. In particular, we will introduce such formulation for fermions and vector fields for the first time.

The manuscript is organised as follows: In section 2, we introduce some of the key concepts of the Wigner formalism and its functional extension in Quantum Field Theory. The next four sections are devoted to find the stationary Wigner functionals of the real and scalar field, the massless electromagnetic field and the fermionic field, respectively. Lastly, we will briefly develop the fundamentals of interacting field theories in phase space, which are necessary in order to calculate transition amplitudes of different processes. The main conclusions are summarised in section 8.

\section{Wigner functional}

The term \textit{Quantum Physics in Phase Space} refers to an alternative representation of Quantum Mechanics, equivalent to the conventional Hilbert space and the path integral formulation. Its main characteristic is the fact that the state of a system is described simultaneously by the positions and momenta of the particles which conform it. Each quantum state is associated to its Wigner function \cite{Wigner_function}:

\begin{equation}\label{Wigner function}
    W(x,p) = \frac{1}{2\pi} \int_{-\infty}^{\infty} dy\ \psi^{*}\left(x - \frac{\hbar}{2}y\right)\, e^{-i p y}\, \psi\left(x + \frac{\hbar}{2}y\right),
\end{equation}

where $\psi(x)$ is the wave function that describes the quantum state in the Schrödinger position representation. In order to construct a non-commutative algebra, the Moyal star product between functions in phase space is introduced as \cite{Groenewold}:

\begin{equation}\label{exp_star}
    f(x,p) \star g(x,p) = f(x,p)\ e^{\frac{i\hbar}{2}(\overleftarrow{\partial_{x}}\overrightarrow{\partial_{p}}-\overleftarrow{\partial_{p}}\overrightarrow{\partial_{x}})} g(x,p).
\end{equation}

Alternatively, this can be expressed in terms of a Bopp shift \cite{Zachos_D58}:

\begin{equation}\label{Bopp_Shift}
    f(x,p) \star g(x,p) = f\left(x + \frac{i\hbar}{2}\partial_{p}, p - \frac{i\hbar}{2}\partial_{x}\right) g(x,p).
\end{equation}

Finally, the dynamics of a quantum state is determined by Moyal's equation \cite{Moyal}:

\begin{equation}\label{Moyal equation}
    \frac{\partial W}{\partial t} = \{\{H,W\}\},
\end{equation}

where the Moyal bracket is defined as $i\hbar\{\{f,g\}\} = f\star g - g \star f$. Thus, stationary states must satisfy the eigenvalue equation

\begin{equation}
    H(x,p)\star W(x,p) = W(x,p)\star H(x,p) = E\, W(x,p).
\end{equation}

In order to extend this formulation to quantum field theory, we will use Jackiw's functional representation \cite{Jackiw,Hatfield}. According to this, a quantum system is described by a Hamiltonian density $\mathcal{H}(\hat{\phi},\hat{\pi})$ which depends on the field operators $\hat{\phi}$ and their canonical conjugate momenta $\hat{\pi}$. Since we are working in the Schrödinger picture, these operators are time independent, $\hat{\phi} = \hat{\phi}(\textbf{x})$, $\hat{\pi} = \hat{\pi}(\textbf{x})$. In the position representation, a basis of eigenstates of the field operators is constructed:

\begin{equation}
    \hat{\phi}|\phi\rangle = \phi(\textbf{x})|\phi\rangle.
\end{equation}

Thus, we can define the functional associated with a given state $\ket{\Psi}$ as $\Psi[\phi] = \langle\phi|\Psi\rangle$. From this, one can apply a functional Wigner transform and obtain the Wigner functional of that state \cite{Zachos_field}:

\begin{align}\label{Wigner transform QFT}
    &W[\phi,\pi] = \int \mathcal{D}\eta \ \Psi^{*}\left[\phi-\frac{\hbar}{2}\eta\right]e^{-i\int d^3\textbf{x}\ \eta(\textbf{x})\pi(\textbf{x})} \Psi\left[\phi+\frac{\hbar}{2}\eta\right],
\end{align}

where the integration measure is given by $\mathcal{D}\eta =\\ \prod_{\textbf{x}}(d\eta(\textbf{x})/2\pi)$. Similarly to the finite-dimensional case, the Moyal star product is introduced as the equivalent of the operator product:

\begin{align}\label{star product QFT}
    &A[\phi,\pi]\star B[\phi,\pi] = A[\phi,\pi]\exp\left(\frac{i\hbar}{2}\int d^{3}\textbf{x}\ \frac{\overleftarrow{\delta}}{\delta\phi(\textbf{x})}\frac{\overrightarrow{\delta}}{\delta\pi(\textbf{x})} - \nonumber \right.\\ & \left. \frac{\overleftarrow{\delta}}{\delta\pi(\textbf{x})}\frac{\overrightarrow{\delta}}{\delta\phi(\textbf{x})} \right) B[\phi,\pi],
\end{align}

where $\delta/\delta\phi(\textbf{x})$ denotes the functional derivative with respect to $\phi(\textbf{x})$. Or, alternatively:

\begin{align}
    &A[\phi,\pi]\star B[\phi,\pi] = A\left[\phi + \frac{i\hbar}{2}\partialfunc{}{\pi(\textbf{x})}, \pi - \frac{i\hbar}{2}\partialfunc{}{\phi(\textbf{x})}\right]B[\phi,\pi] =\nonumber\\ &A[\phi,\pi]B\left[\phi - \frac{i\hbar}{2}\frac{\overleftarrow{\delta}}{\delta\pi(\textbf{x})}, \pi + \frac{i\hbar}{2}\frac{\overleftarrow{\delta}}{\delta\phi(\textbf{x})}\right].
\end{align}

With this definition, the dynamics of a system is completely determined by Moyal's equation:

\begin{equation}
  \{\{ H,W \}\} = \partialder{W}{t}.
\end{equation}

In the case of a stationary state, we have to solve the eigenvalue equation $H[\phi,\pi]\star W[\phi,\pi] = W[\phi,\pi]\star H[\phi,\pi] = E\, W[\phi,\pi]$, which is a functional differential equation.

In this work, we will attempt to solve this eigenvalue equation for some of the main free field theories in order to obtain the Wigner functionals associated with the vacuum and stationary particle states. This will allow us to compare the phase space formulation of quantum field theory to other mainstream approaches. 

\section{Real scalar field}

For the real scalar field, the Hamiltonian is given by:

\begin{equation}
    \Hamiltonian = \int d^3\textbf{x}\ \frac{1}{2}\left[\pi^2(\textbf{x}) + \phi(\textbf{x})(-\nabla^2 + m^2)\phi(\textbf{x})\right].
\end{equation}

The real and imaginary parts of the eigenvalue equation for the Wigner functional are obtained by applying the star product with this Hamiltonian:

\begin{align}\label{Real_scalar_eigenvalues_simple}
    &\frac{1}{2}\int d^{3}\textbf{x} \left[ \pi^2(\textbf{x})- \frac{\hbar^2}{4}\partialfunc{}{\pi(\textbf{x})}(-\nabla^2 + m^2)\partialfunc{}{\pi(\textbf{x})} \right.\nonumber  \\ 
    &\left. + \phi(\textbf{x})(-\nabla^2 + m^2)\phi(\textbf{x}) - \frac{\hbar^2}{4}\partialfunc{^2}{\phi^2(\textbf{x})} \right]\Wigner \nonumber\\
    &= E\, \Wigner\,,
\end{align}

\begin{eqnarray}\label{Scalar_imaginary_part}
    &&\int d^3\textbf{x}\ \left( \pi(\textbf{x})\partialfunc{}{\phi(\textbf{x})} - \frac{1}{2}\phi(\textbf{x})(-\nabla^2 + m^2)\partialfunc{}{\pi(\textbf{x})} - \right.\nonumber \\ 
    &&\left. \frac{1}{2}\partialfunc{}{\pi(\textbf{x})}(-\nabla^2 + m^2)\phi(\textbf{x})  \right)\Wigner = 0\,.
\end{eqnarray}

It is clear that this equation is separable in $\phi$ and $\pi$. However, it is involved the computation of an arbitrary state by solving it. Therefore, only the fundamental state (i.e. the vacuum of the Fock space) will be calculated directly. In this case, we assume that the Wigner functional is strictly positive (this can be justified by noting that the free field theory is simply an infinite set of harmonic oscillators, which means that the Wigner function that describes the fundamental state is positive). Thus, the vacuum state can be written as:

\begin{equation}
    W_0[\phi,\pi] = \mathcal{N}\exp\left[-\mathcal{W}[\phi,\pi]\right].
\end{equation}

Now, the operators present in the eigenvalue equation suggest that the Wigner functional $\mathcal{W}$ is quadratic in $\phi$ and $\pi$. Thus, the second derivatives will contribute to the right-hand member energy term, and the squared first derivatives, to the quadratic terms in $\phi(x)$, $\pi(x)$. With that in mind, it is reasonable to assume that the solution can be written as:

\begin{align}\label{Real_scalar_first_solution}
    \mathcal{W}[\phi, \pi] = \int d^3\textbf{x} d^3\textbf{y}\ &\phi(\textbf{x})g_{\phi}(\textbf{x},\textbf{y})\phi(\textbf{y}) +\nonumber\\ &\pi(\textbf{x})g_{\pi}(\textbf{x},\textbf{y})\pi(\textbf{y}),
\end{align}

with $g_{\phi,\pi}(\textbf{x},\textbf{y})$ being symmetric functions of $\textbf{x},\textbf{y}$. One can argue that this is not the most general form of a quadratic functional: the functions $g(\textbf{x},\textbf{y})$ could be replaced, for example, by a differential operator and integrated only in $\textbf{x}$. However, when doing these calculations, it is useful to interpret the fields $\phi(\textbf{x}),\ \pi(\textbf{x})$ as infinite-dimensional vectors, whose components are the values of said fields at each point in space. Under this approach, the differential operators and functions of two variables would be bilinear forms (i.e., infinite dimensional matrices) whose application on the vectors $\phi, \ \pi$ are written as integrals like the one in (\ref{Real_scalar_first_solution}). Since these quantities are invariant under unitary transformations, it is always possible to find a change of basis, which would take the form of a change of variables in the integral, such that the differential operators are transformed into functions of two variables.

As a last premise, it will be assumed that the functions $g_{\phi,\pi}(\textbf{x},\textbf{y})$ are invariant under translation, i.e.  $g_{\phi,\pi}(\textbf{x},\textbf{y}) = g_{\phi,\pi}(\textbf{x}-\textbf{y})$. Therefore, if we perform a Fourier transform on each of the functions:

\begin{align}\label{Scalar_Fourier}
    &\phi(\textbf{x}) = \int \frac{d^3\textbf{k}}{(2\pi)^{3/2}} \ \tilde{\phi}(\textbf{k})e^{i\textbf{k}\cdot \textbf{x}}, \\ &\pi(\textbf{x}) = \int \frac{d^3\textbf{k}}{(2\pi)^{3/2}} \ \tilde{\pi}(\textbf{k})e^{i\textbf{k}\cdot \textbf{x}}, \\ &g_{\phi,\pi}(\textbf{x}- \textbf{y}) = \int \frac{d^3\textbf{k}}{(2\pi)^{3/2}} \ \tilde{g}_{\phi,\pi}(\textbf{k})e^{i\textbf{k}\cdot (\textbf{x}-\textbf{y})},
\end{align}

then the last two equations take the form:

\begin{align}
    \tilde{g}_{\phi}^2(\textbf{k}) = \frac{1}{(2\pi)^3\hbar^2}\omega_{\textbf{k}}^2, \\ \tilde{g}_{\pi}^2(\textbf{k}) = \frac{1}{(2\pi)^3\hbar^2\omega_{\textbf{k}}^2},
\end{align}

with $\omega_{\textbf{k}} = \sqrt{\textbf{k}^2 + m^2}$. Finally, it follows that:

\begin{align}
    &W[\tilde{\phi},\tilde{\pi}] =\nonumber\\= &\mathcal{N}\exp\left[ -\frac{1}{\hbar}\int d^3\textbf{k}\ \omega_{\textbf{k}}\tilde{\phi}(\textbf{k})\tilde{\phi}(-\textbf{k}) +\frac{\tilde{\pi}(\textbf{k})\tilde{\pi}(-\textbf{k})}{\omega_{\textbf{k}}} \right]. 
\end{align}

Note that this solution also satisfies equation (\ref{Scalar_imaginary_part}). Moreover, this Wigner functional calculated by means of the eigenvalue equation agrees with the Wigner transform of the Schrödinger functional in the position representation, which is given by:

\begin{align}
    &\Psi[\phi] = \mathcal{\tilde{N}} \exp \left[ -\frac{1}{2\hbar}\int d^3\textbf{x}\ \phi(\textbf{x})(-\nabla^2 + m^2)^{1/2}\phi(\textbf{x}) \right] = \nonumber\\&=  \mathcal{\tilde{N}} \exp \left[ -\frac{1}{2\hbar}\int \frac{d^3\textbf{k}}{(2\pi)^3}\ \omega_{\textbf{k}} \phi(\textbf{k})\phi(\textbf{-k}) \right].
\end{align}

Performing the Wigner transform, we obtain a functional integral which has the form of a Gaussian with a linear term. Using the well-known result for a gaussian functional integral \cite{Hatfield}, we arrive at:

\begin{align}
    &W_0[\phi,\pi] \propto \exp \left[-\frac{1}{\hbar}\left(\int d^3\textbf{x}\ \phi(\textbf{x})(-\nabla^2 + m^2)^{1/2}\phi(\textbf{x}) + \right. \right. \nonumber\\ &\left.\left.  \pi(\textbf{x})(-\nabla^2 + m^2)^{-1/2}\pi(\textbf{x}) \right) \right],
\end{align}

which agrees with the result obtained using the eigenvalue equation in phase space. It coincides with the one calculated by Curtright and Zachos in reference \cite{Zachos_field}. To calculate the vacuum energy, we substitute the solution for the functions $g_{\phi,\pi}(\textbf{x}-\textbf{y})$ into equation (\ref{Real_scalar_eigenvalues_simple}):

\begin{align}
    &E_0 = \frac{\hbar^2}{4}\left[\int d^3\textbf{x}\ g_{\phi}(\textbf{x},\textbf{x}) +\right. \nonumber\\ &\left. \int d^3\textbf{x} d^3\textbf{y}\ \delta^{(3)}(\textbf{x}-\textbf{y}) (-\nabla_{\textbf{x}}^2 + m^2)g_{\pi}(\textbf{x},\textbf{y})\right] =\nonumber\\= &\frac{\hbar}{2}\int \frac{d^3\textbf{k}}{(2\pi)^3}\ \omega_{\textbf{k}} \ \delta^{(3)}(\textbf{0}). 
\end{align}

This divergence in the energy of the vacuum state is consistent with other QFT formalisms. Finally, the normalization constant can be obtained by imposing that $\langle 0 |0\rangle = 1$ which, in terms of the Wigner function, means: 

\begin{equation}
    \int \mathcal{D}\phi \mathcal{D}\pi\ W_0[\phi,\pi] = 1,
\end{equation}

where now the integral measures are $\mathcal{D}\phi = \prod_{\textbf{x}}d\phi(\textbf{x})$, $\mathcal{D}\pi = \prod_{\textbf{x}}d\pi(\textbf{x})$. In this way, one arrives at:

\begin{align}
    &1 = \mathcal{N} \frac{(\sqrt{\pi})^{n}}{\sqrt{\det\left(\frac{1}{\hbar}(-\nabla^2 + m^2)^{1/2}\right)}}\times \nonumber\\&\frac{(\sqrt{\pi})^{n}}{\sqrt{\det\left(\frac{1}{\hbar}(-\nabla^2 + m^2)^{-1/2}\right)}} = \mathcal{N}(\hbar\pi)^{n}\,,
\end{align}
with $n \rightarrow \infty$.
In order to check this result, the norm of the vacuum state will be calculated again, this time integrating in Fourier space. Since the Jacobian for this change of variable is $1/(2\pi)^{3/2}$ (note that, apart from this factor, the Fourier transform is unitary, since it preserves the scalar products $\int d^3\textbf{x} f(\textbf{x})g(\textbf{x})$). Therefore, it follows that:

\begin{equation}
\begin{gathered}
    1 = \mathcal{N} \prod_{\textbf{k}}\frac{\sqrt{\hbar\pi}}{\omega_{\textbf{k}}}\prod_{\textbf{k}}\sqrt{\hbar\pi}\omega_{\textbf{k}} \Rightarrow \mathcal{N} = \prod_{\textbf{k}} \frac{1}{\hbar\pi}.
\end{gathered}
\end{equation}

In conclusion, the complete expression of the Wigner functional for the empty state is:

\begin{eqnarray}
   &&W_0[\tilde{\phi},\tilde{\pi}] = \nonumber\\ &&\prod_{\textbf{k}}\frac{1}{\hbar\pi}\exp\left[ -\frac{1}{\hbar}  
   \left( \omega_{\textbf{k}}\tilde{\phi}(\textbf{k})\tilde{\phi}(-\textbf{k}) + \frac{\tilde{\pi}(\textbf{k})\tilde{\pi}(-\textbf{k})}{\omega_{\textbf{k}}} \right)
  \right].
\end{eqnarray}

The next step will be calculating the Wigner functional of an excited state of the field. For this purpose, the creation and annihilation operators are introduced (although, in phase space, they should really be called creation/annihilation functionals):

\begin{align}
    &a^{\dagger}(\textbf{k}) = \frac{1}{\sqrt{2\hbar}} \int d^3\textbf{x}\ \left( \sqrt{\omega_{\textbf{k}}}\phi(\textbf{x}) - \frac{i}{\sqrt{\omega_{\textbf{k}}}}\pi(\textbf{x}) \right)e^{-i\textbf{k}\cdot\textbf{x}}, \nonumber\\
    &a(\textbf{k}) = \frac{1}{\sqrt{2\hbar}} \int d^3\textbf{x}\ \left( \sqrt{\omega_{\textbf{k}}}\phi(\textbf{x}) + \frac{i}{\sqrt{\omega_{\textbf{k}}}}\pi(\textbf{x}) \right)e^{i\textbf{k}\cdot\textbf{x}},
\end{align}

which obey the commutation relations:

\begin{equation}
    \{\{ a(\textbf{k}), a^{\dagger}(\textbf{k}') \}\} = (2\pi)^3\delta^{(3)}(\textbf{k} - \textbf{k}').
\end{equation}

First, we observe that:

\begin{equation}
    a(\textbf{k})\star W_0[\phi,\pi] =  W_0[\phi,\pi] \star a^{\dagger}(\textbf{k}) =0\ \forall \textbf{k},
\end{equation}

which can be checked by using the expression for the functional star product (\ref{star product QFT}) and the explicit form of the vacuum Wigner functional. On the other hand, if the operators are applied in the opposite direction, the resulting functional has the form:

\begin{align}
    &W_1[\phi,\pi] = (2\hbar\omega_{\textbf{k}_1})a^{\dagger}(\textbf{k}_1)\star W_0[\phi,\pi]\star a(\textbf{k}_1) = \nonumber\\ 
    &= 2\omega_{\textbf{k}_1}W_0[\phi,\pi]\left[ \int d^3\textbf{x}\,d^3\textbf{y} \left( \omega_{\textbf{k}_1}\phi(\textbf{x})\phi(\textbf{y}) + \frac{\pi(\textbf{x})\pi(\textbf{y})}{\omega_{\textbf{k}_1}} - \right. \right.\nonumber \\ &\left. \left. i(\phi(\textbf{x})\pi(\textbf{y})-\phi(\textbf{y})\pi(\textbf{x})) - \frac{\hbar}{2}\delta^{(3)}(\textbf{x} - \textbf{y}) \right) e^{i\textbf{k}_1(\textbf{x} - \textbf{y})} \right].
\end{align}

To determine the meaning of these states, one can see that both the Hamiltonian and the linear momentum functionals can be expressed in terms of the creation and annihilation operators as:

\begin{align}
    &H = \int d^3\textbf{x}\ \frac{1}{2}\left[\pi^2(\textbf{x}) + \phi(\textbf{x})(-\nabla^2 + m^2)\phi(\textbf{x})\right] = \nonumber\\ &\frac{\hbar}{2} \int \frac{d^3\textbf{k}}{(2\pi)^3}\ \omega_\textbf{k} \left( a^{\dagger}(\textbf{k})\star a(\textbf{k}) + a(\textbf{k})\star a^{\dagger}(\textbf{k}) \right),
\end{align}
\begin{align}
    &P_j = -\int d^3\textbf{x}\ \phi(\textbf{x})\star\partial_j \pi(\textbf{x})= \nonumber\\ &\frac{\hbar}{2}\int \frac{d^3\textbf{k}}{(2\pi)^3}\ k_j \left( a^{\dagger}(\textbf{k})\star a(\textbf{k}) + a(\textbf{k})\star a^{\dagger}(\textbf{k}) \right).
\end{align}

Expressing these functionals as functions of the ladder ones allows us to easily deduce the following commutation relations:

\begin{equation}
\begin{gathered}
    \{\{ H, a^{\dagger}(\textbf{k}) \}\} = \{\{ a(\textbf{k}), H \}\} = \hbar\omega_{\textbf{k}},  \\ \{\{ P_j, a^{\dagger}(\textbf{k}) \}\} = \{\{ a(\textbf{k}), P_j \}\} = \hbar k_j,
\end{gathered}
\end{equation}

Using them, one arrives at the energy and momentum of the state represented by the excited Wigner functional:

\begin{align}
    &H\star W_1 = W_1 \star H = (E_0 + \hbar\omega_{\textbf{k}_1})W_1 = E_1\, W_1, \nonumber\\
     &P_j\star W_1 = W_1\star P_j = \hbar (k_1)_j\,W_1.
\end{align}

Thus, we can conclude that the functional $W_1[\phi,\pi]$ represents the state of a particle having momentum and energy given by $\textbf{k}_1$, $\omega_{\textbf{k}_1}$. In a completely analogous way, functionals corresponding to states with several particles are constructed by repeatedly applying the creation operator on the vacuum state.

\section{Complex scalar field}

The procedure for the complex scalar field is very similar to the real one. This time, the Hamiltonian is:

\begin{equation}
    H[\phi, \pi] = \int d^3\textbf{x} \ \pi^{\dagger}(\textbf{x})\pi(\textbf{x}) + \phi^{\dagger}(-\nabla^2 + m^2)\phi(\textbf{x}).
\end{equation}

Since the fields are now complex, one must be careful when applying the star product with conjugate fields. If the fields are expressed as $\phi = \phi_1 + i\phi_2$, $\pi = \pi_1 + i\pi_2$, then:

\begin{equation}
\begin{gathered}
    \phi_{1,2}(\textbf{x})\star = \phi_{1,2} + \frac{i\hbar}{2}\partialfunc{}{\pi_{1,2}(\textbf{x})}, \\ \pi_{1,2}(\textbf{x})\star = \pi_{1,2} - \frac{i\hbar}{2}\partialfunc{}{\phi_{1,2}(\textbf{x})}.
\end{gathered}
\end{equation}

Therefore, in terms of the field $\phi$ and its complex conjugate:

\begin{align}
    &\phi(\textbf{x})\star = (\phi_1(\textbf{x})\star) + i(\phi_2(\textbf{x})\star) = \phi(\textbf{x}) + \frac{i\hbar}{2}\partialfunc{}{\pi^{\dagger}(\textbf{x})}, \nonumber\\ &\phi^{\dagger}(\textbf{x})\star = (\phi_1(\textbf{x})\star) - i(\phi_2(\textbf{x})\star) = \phi(\textbf{x}) + \frac{i\hbar}{2}\partialfunc{}{\pi(\textbf{x})}.\\
    &\pi(\textbf{x})\star = (\pi_1(\textbf{x})\star) + i(\pi_2(\textbf{x})\star) = \pi(\textbf{x}) - \frac{i\hbar}{2}\partialfunc{}{\phi^{\dagger}(\textbf{x})}, \nonumber\\ &\pi^{\dagger}(\textbf{x})\star = (\pi_1(\textbf{x})\star) - i(\pi_2(\textbf{x})\star) = \pi(\textbf{x}) - \frac{i\hbar}{2}\partialfunc{}{\phi(\textbf{x})}.
\end{align}

One should not be alarmed because the relations between $\phi$ and $\phi^{\dagger}$ (resp. $\pi$ and $\pi^{\dagger}$) are not complex conjugates of each other. Recall that, when performing the adjoint of a product of operators in Hilbert space, the order of the factors is reversed.

In a completely analogous way to the real scalar field, it can be shown that the solution for the ground state is given by a functional quadratic exponential:

\begin{align}
    W_0[\tilde{\phi}, \tilde{\pi}] \propto \exp\left[ -\frac{2}{\hbar}\int d^3\textbf{k}\ \tilde{\phi}^{\dagger}(\textbf{k})\omega_{\textbf{k}}\tilde{\phi}(\textbf{k}) + \frac{\tilde{\pi}^{\dagger}(\textbf{k})\tilde{\pi}(\textbf{k})}{\omega_{\textbf{k}}}  \right].
\end{align}

In this case, we have to generate two sets of creation/\\annihilation functionals.

\begin{align}
    &a^{\dagger}(\textbf{k}) = \frac{1}{\sqrt{2\hbar}} \int d^3\textbf{x}\ \left( \sqrt{\omega_{\textbf{k}}}\phi^{\dagger}(\textbf{x}) - \frac{i}{\sqrt{\omega_{\textbf{k}}}}\pi(\textbf{x}) \right)e^{-i\textbf{k}\cdot\textbf{x}},\nonumber\\
    &a(\textbf{k}) = \frac{1}{\sqrt{2\hbar}} \int d^3\textbf{x}\ \left( \sqrt{\omega_{\textbf{k}}}\phi(\textbf{x}) + \frac{i}{\sqrt{\omega_{\textbf{k}}}}\pi^{\dagger}(\textbf{x}) \right)e^{i\textbf{k}\cdot\textbf{x}},\nonumber\\
    &b^{\dagger}(\textbf{k}) = \frac{1}{\sqrt{2\hbar}} \int d^3\textbf{x}\ \left( \sqrt{\omega_{\textbf{k}}}\phi(\textbf{x}) - \frac{i}{\sqrt{\omega_{\textbf{k}}}}\pi^{\dagger}(\textbf{x}) \right)e^{-i\textbf{k}\cdot\textbf{x}},\nonumber\\
    &b(\textbf{k}) = \frac{1}{\sqrt{2\hbar}} \int d^3\textbf{x}\ \left( \sqrt{\omega_{\textbf{k}}}\phi^{\dagger}(\textbf{x}) + \frac{i}{\sqrt{\omega_{\textbf{k}}}}\pi(\textbf{x}) \right)e^{i\textbf{k}\cdot\textbf{x}},
\end{align}

which satisfy the commutation relations:

\begin{align}
    &\{\{ a(\textbf{k}),a^{\dagger}(\textbf{k}') \}\} =\{\{ b(\textbf{k}),b^{\dagger}(\textbf{k}') \}\} = \\ & =(2\pi)^3\delta^{(3)}(\textbf{k} - \textbf{k}'), \nonumber
\end{align}

with the rest of them being zero. As in the real case, it is found that the Hamiltonian and the linear momentum can be expressed as:

\begin{align}
    &H = \frac{\hbar}{2}\int \frac{d^3\textbf{k}}{(2\pi)^3}\omega_{\textbf{k}}(a^{\dagger}(\textbf{k})\star a(\textbf{k}) + a(\textbf{k})\star a^{\dagger}(\textbf{k}) +\nonumber\\ &b^{\dagger}(\textbf{k})\star b(\textbf{k}) + b(\textbf{k})\star b^{\dagger}(\textbf{k})),
\end{align}
\begin{align}
    &P_j =  \frac{\hbar}{2}\int \frac{d^3\textbf{k}}{(2\pi)^3}k_j(a^{\dagger}(\textbf{k})\star a(\textbf{k}) + a(\textbf{k})\star a^{\dagger}(\textbf{k}) + \nonumber\\ &b^{\dagger}(\textbf{k})\star b(\textbf{k}) + b(\textbf{k})\star b^{\dagger}(\textbf{k})).
\end{align}

Therefore, the states $a^{\dagger}(\textbf{k})\star W_0 \star a(\textbf{k})$ and $b^{\dagger}(\textbf{k})\star W_0 \star b(\textbf{k})$ represent states of a particle (antiparticle) with momentum $\textbf{k}$ and energy $\omega_{\textbf{k}}$. Finally, we consider the quantity:

\begin{equation}\label{scalar charge}
    Q = i\int d^3\textbf{x}\ \phi^{\dagger}(\textbf{x})\pi^{\dagger}(\textbf{x}) - \phi(\textbf{x})\pi(\textbf{x}),
\end{equation}

which will be denoted as charge. In terms of the ladder functionals, it has the form:

\begin{equation}
    Q = \hbar\int \frac{d^3\textbf{k}}{(2\pi)^3} \ (a^{\dagger}(\textbf{k})\star a(\textbf{k}) - b(\textbf{k})\star b^{\dagger}(\textbf{k})).
\end{equation}

Although this has not been done for the previous functionals, in this case we will make use of normal ordering of the creation and annihilation functionals in order to avoid the vacuum state having a non-zero charge. This can be justified due to the fact that there is some ambiguity in the ordering of the fields in the definition of the charge (\ref{scalar charge}). Therefore, the expression to be used is:

\begin{equation}
    :Q: = \hbar\int \frac{d^3\textbf{k}}{(2\pi)^3} \ (a^{\dagger}(\textbf{k})\star a(\textbf{k}) - b^{\dagger}(\textbf{k})\star b(\textbf{k})).
\end{equation}

It is easy to show that the charge $Q$ and the Hamiltonian $H$ commute under Moyal's bracket. Thus, using Moyal's equation, it follows that $Q$ is conserved:

\begin{equation}
    \der{Q}{t} = \{\{ Q, H \}\} = 0.
\end{equation}

Consequently, the states obtained by consecutively applying the ladder functionals are eigenstates of the charge functional $Q$, with eigenvalue $n_a - n_b$, where $n_a$ (resp. $n_b$) is the number of times we have applied the functional $a^{\dagger}\star()\star a$ (resp. $b^{\dagger}\star()\star b$) to the vacuum state. Therefore, they can be interpreted as states with $n_a$ particles with charge +1 and $n_b$ antiparticles with charge -1.

\section{Massless electromagnetic field}

To construct the Wigner functionals of the free electromagnetic field, it is necessary to make two new considerations with respect to the scalar case: firstly, the fields have more than one component and, secondly, they have certain degrees of freedom under gauge transformations. 
With this in mind, the configuration space will be made up by the time-independent $A_{\mu}(\textbf{x})$ fields (recall that we are working in the Schrödinger picture, where the Hilbert space operators do not evolve in time). Because of this, it is very useful to choose the temporal gauge, $A_0 = 0$. This way, the conjugate momentum of the vector potential, that is, the electric field $\textbf{E}(\textbf{x})$, will be completely independent of $\textbf{A}(\textbf{x})$ (which is essential when working in phase space). However, there is still some residual gauge freedom left:

\begin{equation}
    \textbf{A}(\textbf{x}) \rightarrow \textbf{A}(\textbf{x}) - \nabla \Lambda(\textbf{x}).
\end{equation}

We could impose some other condition on the fields, for example, the Coulomb gauge $\nabla\cdot\textbf{A}(\textbf{x}) = 0$. However, this implies that we would have to modify the canonical commutation rules to meet this condition, replacing the conventional Dirac delta by a transverse Dirac delta. Since we do not want to change the definition of the star product, we will choose not to impose any further conditions on the fields.

On the other hand, the fact that the fields are time-independent and that $A_0 = 0$ eliminates the presence of Gauss's law $\nabla\cdot \textbf{E}(\textbf{x}) = 0$. To solve this, following the example of Hatfield \cite{Hatfield}, the following constraint will be added as a condition on the states:

\begin{align}\label{Gauss contrain}
    &\nabla\cdot\textbf{E}(\textbf{x})\star W[\textbf{A},\textbf{E}] = W[\textbf{A},\textbf{E}] \star \nabla\cdot\textbf{E}(\textbf{x}) = 0 \Rightarrow \nonumber\\ &\nabla\cdot \partialfunc{W}{\textbf{A}(\textbf{x})} = 0, \ \nabla\cdot \textbf{E}(\textbf{x}) = 0.
\end{align}

It can be seen that this guarantees that the Wigner functional is invariant under residual gauge transformations. Indeed, if $\delta\textbf{A}(\textbf{x}) = -\nabla\Lambda(\textbf{x})$, then

\begin{align}
    &\delta W[\textbf{A},\textbf{E}] = \int d^3\textbf{x}\ \partialfunc{W}{\textbf{A}(\textbf{x})}\cdot \delta \textbf{A}(\textbf{x}) = \nonumber\\ &-\int d^3\textbf{x}\ \partialfunc{W}{\textbf{A}(\textbf{x})}\cdot \nabla \Lambda(\textbf{x}) = 0,
\end{align}

where we have integrated by parts. Equivalently, one can interpret that the action of $\partialfunc{}{\textbf{A}(\textbf{x})}$ on the Wigner functional 'extracts' the factors $\textbf{A}(\textbf{x})$ appearing in \\$W[\textbf{A},\textbf{E}]$. Thus, the condition (\ref{Gauss contrain}) implies that all components of $\textbf{A}(\textbf{x})$ that are present in the Wigner functional are transverse, while the longitudinal ones are eliminated. Since the transverse components $\textbf{A}_T$ are gauge invariant (while the longitudinal components $\textbf{A}_L$ are not) it follows that $W[\textbf{A},\textbf{E}]$ is also gauge invariant.
On the other hand, one should not forget the other condition that we imposed, that is, $\nabla\cdot \textbf{E}(\textbf{x}) = 0$. In contrast to the vector potential, where this requisite is defined by its action on the functionals, the electric field is directly restricted, i.e. only the transverse components of the electric field are considered. We will have to take this into account at all times when calculating Wigner functionals, since it will not be denoted explicitly.
One problem with this choice of gauge is that, when calculating expected values, unwanted divergences appear. This is because any physical quantity $\mathcal{O}[\textbf{A},\textbf{E}]$ must be gauge invariant, which implies that it only depends, like the Wigner functional, on the transverse components of the vector potential. Therefore, the expectation value:

\begin{equation}\label{Expected value EM}
    \langle \mathcal{O} \rangle = \frac{\int \mathcal{D}\textbf{A}\mathcal{D}\textbf{E}\ \mathcal{O}[\textbf{A}_T,\textbf{E}]W[\textbf{A}_T,\textbf{E}]}{\int \mathcal{D}\textbf{A}\mathcal{D}\textbf{E}\ W[\textbf{A}_T,\textbf{E}]},
\end{equation}

adds the values of $\mathcal{O}[\textbf{A}_T,\textbf{E}_T]$ and $W[\textbf{A}_T,\textbf{E}_T]$ infinitely many times, since the integration measures can be factored into $\mathcal{D}\textbf{A} = \mathcal{D}\textbf{A}_T \mathcal{D}\textbf{A}_L$ (this does not apply to the electric field because, by definition, all possible values of the field are transverse). However, for abelian gauge theories such as the massless electromagnetic field \\(where the group of gauge transformations is $U(1)$), the longitudinal and transverse components are easy to separate, so that their contributions in the numerator and denominator of (\ref{Expected value EM}) cancel.
Once all these details have been taken into account, one can proceed to calculate the Wigner functionals. In the temporal gauge, the Hamiltonian of the theory is given by:

\begin{align}
    &H = \frac{1}{2}\int d^3\textbf{x}\ \textbf{E}^2(\textbf{x}) + \textbf{B}^2(\textbf{x}) =\nonumber\\ &\int d^3\textbf{x}\ \textbf{E}^2(\textbf{x}) + (\nabla\times\textbf{A}(\textbf{x}))^2.
\end{align}

This leads, again, to an eigenvalue equation imposed on the stationary states. In order to calculate the vacuum Wigner functional, this equation will be expressed directly in the Fourier space of the fields, i.e:

\begin{align}
    &\textbf{A}(\textbf{x}) = \int \frac{d^3\textbf{k}}{(2\pi)^{3/2}}\ \tilde{\textbf{A}}(\textbf{k})e^{i\textbf{k}\cdot\textbf{x}}, \nonumber\\ 
    &\textbf{E}(\textbf{x}) = \int \frac{d^3\textbf{k}}{(2\pi)^{3/2}}\ \tilde{\textbf{E}}(\textbf{k})e^{i\textbf{k}\cdot\textbf{x}},\nonumber\\
    &\partialfunc{}{\textbf{A}(\textbf{x})} = \int \frac{d^3\textbf{k}}{(2\pi)^{3/2}}\ \partialfunc{}{\tilde{\textbf{A}}(\textbf{k})}e^{-i\textbf{k}\cdot\textbf{x}}, \nonumber\\ 
    &\partialfunc{}{\textbf{E}(\textbf{x})} = \int \frac{d^3\textbf{k}}{(2\pi)^{3/2}}\ \partialfunc{}{\tilde{\textbf{E}}(\textbf{k})}e^{-i\textbf{k}\cdot\textbf{x}},
\end{align}

where the multicomponent functional Dirac delta acts in the following way:

\begin{align}
        &\partialfunc{\tilde{A_i}(\textbf{k})}{\tilde{A_j}(\textbf{k}')} = \delta_{ij} \delta^{(3)}(\textbf{k} - \textbf{k}'), \nonumber\\ &\partialfunc{\tilde{E_i}(\textbf{k})}{\tilde{E_j}(\textbf{k}')} = \delta_{ij} \delta^{(3)}(\textbf{k} - \textbf{k}').
\end{align}

The condition that the electric field is purely transverse translates into $\textbf{k}\cdot\tilde{\textbf{E}}(\textbf{k}) = 0\ \forall \textbf{k}$. Based on these expressions, it can be seen that the equation that determines the vacuum state takes the form:

\begin{align}
    &\frac{1}{2}\int d^3 \textbf{k}\ \left[ \tilde{\textbf{E}}(\textbf{k})\cdot\tilde{\textbf{E}}(-\textbf{k}) - \frac{\hbar^2}{4}\left(\textbf{k}\times \partialfunc{}{\tilde{\textbf{E}}(\textbf{k})}\right)\cdot \right.\nonumber\\
    &\left. \left(\textbf{k}\times \partialfunc{}{\tilde{\textbf{E}}(-\textbf{k})}\right)  +(\textbf{k}\times\tilde{\textbf{A}}(\textbf{k}))\cdot(\textbf{k}\times\tilde{\textbf{A}}(-\textbf{k})) - \right.\nonumber\\
    &\left. \frac{\hbar^2}{4}\partialfunc{}{\tilde{\textbf{A}}(\textbf{k})}\cdot\partialfunc{}{\tilde{\textbf{A}}(-\textbf{k})}\right] W_0[\tilde{\textbf{A}},\tilde{\textbf{E}}] = E_0\, W_0[\tilde{\textbf{A}},\tilde{\textbf{E}}].
\end{align}

As in the previous cases, we assume that the Wigner functional that represents the fundamental state is an exponential of the fields. If $\mathcal{W}$ is taken as a quadratic functional in $\textbf{A}, \, \textbf{E}$, as suggested by the presence of the functional derivatives and the quadratic terms and constants, the equation splits into two parts:

\begin{align}
       &\frac{\hbar^2}{8}\int d^3\textbf{k}\ \left[ -\left( \textbf{k}\cdot\partialfunc{}{\tilde{\textbf{E}}(\textbf{k})} \right)\left( \textbf{k}\cdot\partialfunc{}{\tilde{\textbf{E}}(-\textbf{k})} \right)\mathcal{W} \right. \nonumber\\ &\left. +\textbf{k}^2\frac{\delta^2 \mathcal{W}}{\delta\tilde{\textbf{E}}(\textbf{k})\cdot\delta\tilde{\textbf{E}}(-\textbf{k})} + \frac{\delta^2\mathcal{W}}{\delta\tilde{\textbf{A}}(\textbf{k})\delta\tilde{\textbf{A}}(-\textbf{k})} \right] = E_0,
\end{align}
\begin{align}
       &\frac{\hbar^2}{4}\int d^3\textbf{k} \ \left(\textbf{k}\times \partialfunc{\mathcal{W}}{\tilde{\textbf{E}}(\textbf{k})}\right)\cdot\left(\textbf{k}\times \partialfunc{\mathcal{W}}{\tilde{\textbf{E}}(-\textbf{k})}\right) +\nonumber\\ &\partialfunc{\mathcal{W}}{\tilde{\textbf{A}}(\textbf{k})}\cdot\partialfunc{\mathcal{W}}{\tilde{\textbf{A}}(-\textbf{k})} = \int d^3\textbf{k}\ \tilde{\textbf{E}}(\textbf{k})\cdot \tilde{\textbf{E}}(-\textbf{k}) + \nonumber\\ &(\textbf{k}\times\tilde{\textbf{A}}(\textbf{k}))\cdot(\textbf{k}\times\tilde{\textbf{A}}(-\textbf{k})),
\end{align}

where the functional Laplacian is defined as:

\begin{equation}
    \frac{\delta^2\mathcal{W}}{\delta\tilde{\textbf{E}}(\textbf{k})\cdot\delta\tilde{\textbf{E}}(-\textbf{k})} = \sum_i \frac{\delta^2\mathcal{W}}{\delta\tilde{E_i}(\textbf{k})\delta\tilde{E_i}(-\textbf{k})}\,.
\end{equation}

Contrary to the previous fields, a solution of the form:

\begin{align}
    &\mathcal{W}[\tilde{\textbf{A}},\tilde{\textbf{E}}] = \nonumber\\&\int d^3\textbf{k}\ \tilde{g}_{\textbf{A}}(\textbf{k})\tilde{\textbf{A}}(\textbf{k})\cdot \tilde{\textbf{A}}(-\textbf{k}) + \tilde{g}_{\textbf{E}}(\textbf{k})\tilde{\textbf{E}}(\textbf{k})\cdot \tilde{\textbf{E}}(-\textbf{k}), 
\end{align}

doesn't satisfy the Gauss condition (\ref{Gauss contrain}) which, in Fourier space, can be expressed as $\textbf{k}\cdot \delta W/\delta\tilde{\textbf{A}}(\textbf{k}) = 0$. Since, when applying these operators on the Wigner vacuum functional, terms proportional to $\textbf{k}\cdot \tilde{\textbf{A}}(\textbf{k})$ appear, one can guarantee (\ref{Gauss contrain}) if the following substitution is made in the initial quadratic expression (see Hatfield \cite{Hatfield}):

\begin{equation}
    \tilde{\textbf{A}}(\textbf{k}) \rightarrow \tilde{\textbf{A}}(\textbf{k}) - \frac{(\textbf{k}\cdot\tilde{\textbf{A}}(\textbf{k}))}{\textbf{k}^2} = \tilde{\textbf{A}}_T(\textbf{k}).
\end{equation}

It is easy to verify that the restrictions imposed on the states are now fulfilled. In order to calculate the functions $\tilde{g}_{\textbf{A},\textbf{E}}$, we return to the eigenvalue equations. Assuming that $g_{\textbf{A},\textbf{E}}(\textbf{x})$ are real functions, that is,  $\tilde{g}_{\textbf{A},\textbf{E}}(\textbf{k}) = \tilde{g}_{\textbf{A},\textbf{E}}(-\textbf{k})$, and taking into account the vector relation $(\textbf{a}\times\textbf{b})\cdot (\textbf{c}\times\textbf{d}) = (\textbf{a}\cdot\textbf{c})(\textbf{b}\cdot\textbf{d}) - (\textbf{a}\cdot\textbf{d})(\textbf{b}\cdot\textbf{c})$, it follows that the functions $\tilde{g}_{\textbf{A},\textbf{E}}(\textbf{k})$ must satisfy:

\begin{equation}
    \hbar^2 \tilde{g}^2_\textbf{E}(\textbf{k})\textbf{k}^2 = 1, \quad \hbar^2 \tilde{g}^2_\textbf{A}(\textbf{k}) = \textbf{k}^2.
\end{equation}

The final expression of the Wigner functional for the fundamental state of the electromagnetic field is:

\begin{align}\label{Wigner functional EM}
    &W[\tilde{\textbf{A}},\tilde{\textbf{E}}] =\nonumber\\ &\mathcal{N}\prod_{\textbf{k}}\exp\left[-\left(\omega_{k}\tilde{\textbf{A}}_T(\textbf{k})\cdot\tilde{\textbf{A}}_T(-\textbf{k}) + \frac{\tilde{\textbf{E}}(\textbf{k})\cdot\tilde{\textbf{E}}(-\textbf{k})}{\omega_{\textbf{k}}} \right) \right], 
\end{align}

where $\omega_{\textbf{k}} = |\textbf{k}|$. We emphasize again that $\tilde{\textbf{E}}(\textbf{k})$ fields are transverse by definition, which is why it is not explicitly stated.
One can directly check that this expression coincides with the Wigner transform of the Schrödinger functional, which is given by \cite{Hatfield}:

\begin{equation}
    \Psi[\tilde{\textbf{A}}] = \tilde{\mathcal{N}}\exp\left[-\frac{1}{2\hbar}\int d^3\textbf{k}\ \omega_{k}\tilde{\textbf{A}}_T(\textbf{k})\cdot\tilde{\textbf{A}}_T(-\textbf{k}) \right].
\end{equation}

If we use this expression in the formula for the Wigner transform (\ref{Wigner transform QFT}), and integrate only over the transverse components of the vector potential (otherwise, there would be an infinite factor in front of the integral), we obtain the exact same result as with the eigenvalue equation.

Now, in order to determine the energy of the vacuum state, the second functional derivatives of the solution are calculated:

\begin{align}
    &\frac{\delta^2\mathcal{W}}{\delta\tilde{E}_i(\textbf{k})\delta\tilde{E}_i(-\textbf{k})} = \frac{2}{\omega_{\textbf{k}}\hbar}\delta^{(3)}(\textbf{0}),\nonumber \\ &\frac{\delta^2\mathcal{W}}{\delta\tilde{A}_i(\textbf{k})\delta\tilde{A}_i(-\textbf{k})} = \frac{2\omega_{\textbf{k}}}{\hbar}\delta^{(3)}(\textbf{0})\left( 1 - \frac{k_i^2}{\textbf{k}^2} \right).
\end{align}

Recalling that the vectors $\tilde{\textbf{E}}(\textbf{k})$, which are transverse, have only two nonzero components, we arrive at:

\begin{align}
    &E_0 = \frac{\hbar^2}{8}\int d^3\textbf{k}\ \left[ -\left( \textbf{k}\cdot\partialfunc{}{\tilde{\textbf{E}}(\textbf{k})} \right)\left( \textbf{k}\cdot\partialfunc{}{\tilde{\textbf{E}}(-\textbf{k})} \right)\mathcal{W} + \right. \nonumber\\ &\left. \textbf{k}^2\frac{\delta^2 \mathcal{W}}{\delta\tilde{\textbf{E}}(\textbf{k})\cdot\delta\tilde{\textbf{E}}(-\textbf{k})} + \frac{\delta^2\mathcal{W}}{\delta\tilde{\textbf{A}}(\textbf{k})\delta\tilde{\textbf{A}}(-\textbf{k})} \right] = \nonumber\\
    & =\int d^3\textbf{k}\ \hbar\omega_{\textbf{k}}\delta^{(3)}(\textbf{0}).
\end{align}

Next, the creation and annihilation functionals will be constructed as:

\begin{align}\label{Ladder EM}
    &a^{\dagger}(\textbf{k},\lambda) = \frac{1}{\sqrt{2\hbar}}\int d^3\textbf{x} \ \bm{\epsilon}(\textbf{k},\lambda)\cdot\nonumber\\ &\left( \sqrt{\omega_{\textbf{k}}}\textbf{A}(\textbf{x}) - \frac{i}{\sqrt{\omega_{\textbf{k}}}}\textbf{E}(\textbf{x}) \right)e^{-i\textbf{k}\cdot \textbf{x}}\nonumber\\
    &a(\textbf{k},\lambda) = \frac{1}{\sqrt{2\hbar}}\int d^3\textbf{x} \ \bm{\epsilon}^{*}(\textbf{k},\lambda)\cdot\nonumber\\ &\left( \sqrt{\omega_{\textbf{k}}}\textbf{A}(\textbf{x}) + \frac{i}{\sqrt{\omega_{\textbf{k}}}}\textbf{E}(\textbf{x}) \right)e^{i\textbf{k}\cdot \textbf{x}},
\end{align}

where $\bm{\epsilon}(\textbf{k},\lambda)$ are orthonormal vectors (although they need not be real) associated with each transverse polarization ($\lambda = 1,2$), that is, $\bm{\epsilon}(\textbf{k},\lambda)\cdot\textbf{k} = 0$. Actually, the choice of the temporal gauge does not prevent us from defining longitudinal creation and annihilation functionals corresponding to $\bm{\epsilon}(\textbf{k},\lambda)\| \textbf{k}$. However, it can be easily checked that any state formed from the action of these longitudinal functionals would not satisfy the condition (\ref{Gauss contrain}). Hence, it makes no sense to define them.

Consequently, the functional corresponding to a photon with momentum $\textbf{k}_1$ and polarization $\lambda$ is calculated as:

\begin{equation}\label{Photon k1}
\begin{split}
    &W_1[\tilde{\textbf{A}},\tilde{\textbf{E}}] = a^{\dagger}(\textbf{k}_1,\lambda)\star W_0[\textbf{A},\textbf{E}]\star a(\textbf{k}_1,\lambda) =\\ &2(2\pi)^3\omega_{\textbf{k}_1}\Big[ \omega_{\textbf{k}_1}(\bm{\epsilon}(\textbf{k}_1,\lambda)\cdot\tilde{\textbf{A}}_T(\textbf{k}_1))(\bm{\epsilon}^*(\textbf{k}_1,\lambda)\cdot\tilde{\textbf{A}}_T(-\textbf{k}_1)) + \\  &\frac{1}{\omega_{\textbf{k}_1}}(\bm{\epsilon}(\textbf{k}_1,\lambda)\cdot\tilde{\textbf{E}}(\textbf{k}_1))(\bm{\epsilon}^*(\textbf{k}_1,\lambda)\cdot\tilde{\textbf{E}}(-\textbf{k}_1))-  \frac{\hbar}{2}\delta^{(3)}(\textbf{0}) +\\
    & i\left( (\bm{\epsilon}(\textbf{k}_1,\lambda)\cdot\tilde{\textbf{A}}_T(\textbf{k}_1))(\bm{\epsilon}^*(\textbf{k}_1,\lambda)\cdot\tilde{\textbf{E}}(-\textbf{k}_1)) - \right.  \\ & \left. (\bm{\epsilon}(\textbf{k}_1,\lambda)\cdot\tilde{\textbf{E}}(\textbf{k}_1))(\bm{\epsilon}^{*}(\textbf{k}_1,\lambda)\cdot\tilde{\textbf{A}}_T(-\textbf{k}_1)) \right) \Big]W_0[\tilde{\textbf{A}},\tilde{\textbf{E}}].
\end{split}
\end{equation}

To show that this state represents a photon of momentum $\textbf{k}_1$ and energy $\omega_{\textbf{k}_1}$, it is noted that the Hamiltonian and the momentum operator can be written as a function of $a^{\dagger}$ and $a$ as:

\begin{align}
        &H = \frac{1}{2}\int d^3\textbf{x}\ \textbf{E}^2(\textbf{x}) +(\nabla\times\textbf{A}(\textbf{x}))^2 =\nonumber\\ &\frac{\hbar}{2}\int \frac{d^3\textbf{k}}{(2\pi)^3}\ \omega_{k} \sum_{\lambda} a^{\dagger}(\textbf{k},\lambda)\star a(\textbf{k},\lambda) + a(\textbf{k},\lambda)\star a^{\dagger}(\textbf{k},\lambda) \\
        &P_j = \int d^3\textbf{x}\ \textbf{E}(\textbf{x})\times(\nabla\times\textbf{A}(\textbf{x})) =\nonumber\\ &\frac{\hbar}{2}\int \frac{d^3\textbf{k}}{(2\pi)^3}\ k_j \sum_{\lambda} a^{\dagger}(\textbf{k},\lambda)\star a(\textbf{k},\lambda) + a(\textbf{k},\lambda)\star a^{\dagger}(\textbf{k},\lambda), 
\end{align}

where we have taken into account the completeness relation of the set of vectors $\left\{\bm{\epsilon}(\textbf{k},1),\bm{\epsilon}(\textbf{k},2),\textbf{k}\right\}$:

\begin{equation}
    \sum_{\lambda}\epsilon_i(\textbf{k},\lambda)\epsilon_j^*(\textbf{k},\lambda) = \delta_{ij} - \frac{k_ik_j}{\textbf{k}^2},
\end{equation}

Therefore, in a completely analogous way as in the previous cases, we see that the functional $W_1[\textbf{A},\textbf{E}]$ defined in (\ref{Photon k1}) corresponds to a photon with momentum $\textbf{k}_1$ and energy $\omega_{\textbf{k}_1} = |\textbf{k}|$.

A final important quantity when analyzing the electromagnetic field is helicity:

\begin{equation}
    h = \frac{\textbf{S}\cdot\textbf{P}}{|\textbf{P}|},
\end{equation}

where the spin $\textbf{S}$ is given by $\textbf{S} = \int d^3\textbf{x}\ \textbf{E}(\textbf{x})\times\textbf{A}(\textbf{x})$. Expressing this functional in terms of the creation and annihilation functionals is an algebraically arduous task, mainly because the invariance under changes of basis of the polarization vectors is not as trivial as in the case of the Hamiltonian or the momentum. However, if one expresses the fields $\textbf{A}$ and $\textbf{E}$ as a function of the creation and annihilation functionals by inverting the relations (\ref{Ladder EM}), it is not difficult to show that linearly polarized states (i.e. states which are created by applying ladder functionals associated with real unit vectors $\bm{\epsilon}(\textbf{k})\bot \textbf{k}$) are not eigenstates of the helicity functional. However, circularly polarized states,

\begin{equation}
    \begin{split}
        W_{\pm} = a^{\dagger}(\textbf{k},\pm)\star W_0\star a(\textbf{k},\pm),
    \end{split}
\end{equation}

where $a^{\dagger}(\textbf{k},\pm) = \frac{1}{\sqrt{2}}(a^{\dagger}(\textbf{k},1)\pm i a^{\dagger}(\textbf{k},2))$, y $a(\textbf{k},\pm) = \frac{1}{\sqrt{2}}(a(\textbf{k},1)\mp i a(\textbf{k},2))$, can be shown to be eigenstates of $h$ with eigenvalue $\pm 1$. In conclusion, one can interpret the state $W_{\pm}$ as a photon with momentum, energy and helicity given by $\textbf{k}$, $\omega_{\textbf{k}}$ and $h = \pm 1$.

\section{Free fermionic field}

The Wigner formalism for fermionic field presents an obvious added difficulty, namely, the canonical commutation rules that characterized the scalar and electromagnetic fields become anti-commutation relations. In the Fock space formulation, this doesn't present a problem, since we simply postulated these relations on the field operators and perform the calculations in the normal way. However, when working with functionals, it is necessary to redefine the nature of the variables and the expression of the star product.

This is immediately apparent when imposing the anti-commutation relations of the field operators. Since $i\hat{\psi}^{\dagger}$ is the conjugate momentum of $\hat{\psi}$, it follows that:

\begin{align}\label{anti-commutation}
    &\{\hat{\psi}_a(\textbf{x}),\hat{\psi}_b(\textbf{y})\} =  \{\hat{\psi}^{\dagger}_a(\textbf{x}),\hat{\psi}^{\dagger}_b(\textbf{y})\} = 0, \nonumber\\ &\{\hat{\psi}_a(\textbf{x}),\hat{\psi}^{\dagger}_b(\textbf{y})\} = \hbar\delta_{ab}\delta^{(3)}(\textbf{x}-\textbf{y}).
\end{align}

Consequently, the eigenvalues of such operators, which are functions of the position $\textbf{x}$, must also satisfy them. Defining the Moyal anti-bracket as:

\begin{align}
    &\{\{ A[\psi,\psi^{\dagger}],B[\psi,\psi^{\dagger}]\}\}_{anti} =\nonumber\\ &A[\psi,\psi^{\dagger}]\star B[\psi,\psi^{\dagger}] + B[\psi,\psi^{\dagger}]\star A[\psi,\psi^{\dagger}],
\end{align}

it follows that $\psi$ and $\psi^{\dagger}$ obey the relations (\ref{anti-commutation}) under the Moyal anti-bracket. In order to satisfy them, we have to redefine the action of the star product:

\begin{align}
    &\psi_a(\textbf{x})\star = \psi_a(\textbf{x}) + \frac{\hbar}{2}\partialfunc{}{\psi^{\dagger}_a(\textbf{x})}, \nonumber\\ &\psi^{\dagger}_a(\textbf{x})\star = \psi^{\dagger}_a(\textbf{x}) + \frac{\hbar}{2}\partialfunc{}{\psi_a(\textbf{x})}.
\end{align}

However, this is not enough. If, with this definition, the first Moyal anti-bracket is calculated, one arrives at:

\begin{align}
    &\psi_a(\textbf{x})\psi_b(\textbf{y}) + \psi_b(\textbf{y})\psi_a(\textbf{x}) = 0,\nonumber\\
    &\psi^{\dagger}_a(\textbf{x})\psi^{\dagger}_b(\textbf{y}) + \psi^{\dagger}_b(\textbf{y})\psi^{\dagger}_a(\textbf{x}) = 0.
\end{align}

From this, we conclude that the functions $\psi(\textbf{x})$ and $\psi^{\dagger}(\textbf{x})$ cannot be usual variables (i.e., c-numbers that commute). Instead, they must be interpreted as Grassmann variables which, by definition, obey anticommutation relations. That is, if $\theta_1\ , \theta_2$ are two Grassmann variables, then $\theta_1\theta_2 + \theta_2\theta_1 = 0$. Therefore, from now on all calculations will be performed taking into account the properties of said type of variables, some of which can be found in Peskin and Schroeder's book \cite{Peskin}.

The fact that the fields anti-commute allows us to define the fermionic functional in phase space. Indeed, if the eigenvalues of the field operators were c-numbers, then $\psi^{\dagger}(\textbf{x})$ would correspond to the adjoint spinor of $\psi(\textbf{x})$. Therefore, the conjugate momentum of the field would be completely determined by the field itself. This would render the definition of phase space meaningless, since the two variables are not independent. However, since the complex conjugate operation is not completely defined for the Grassmann variables (more precisely, it is defined as any antilinear and idempotent application), the conjugate momenta are not determined by the fields and, consequently, can be treated as independent. 

The Hamiltonian of fermionic field theory is the so-called Dirac Hamiltonian:
 
\begin{equation}\label{Dirac Hamiltonian}
    H = \int d^3\textbf{x} \ \psi^{\dagger}\star(\textbf{x})(-i\bm{\alpha}\cdot\nabla + \beta m)\psi(\textbf{x}),
\end{equation}

where the matrices $\alpha_i$ ($i = 1,2,3)$ and $\beta$ can be taken as $\beta = \gamma_0$, $\alpha_i = \gamma_0\gamma_i$, with $\gamma_{\mu}$ ($\mu = 0,1,2,3$) being the Dirac matrices. In order to solve the eigenvalue equation, we will work directly in the Fourier space of the fields, i.e. we will express the Hamiltonian as a function of the eigenvalues of the creation and destruction operators (which, as we mentioned above, are simply functionals of the fields and their momenta in phase space) to avoid having to work with the Dirac matrices. Thus, the fields will be written as:

\begin{align}
    &\psi_{a}(\textbf{x}) = \sum_{s=1}^4 \int \frac{d^3\textbf{p}}{(2\pi)^{3/2}}\  \frac{1}{\sqrt{\omega_{\textbf{p}}}} b(\textbf{p},s)u_a(\textbf{p},s)e^{i\textbf{p}\cdot\textbf{x}}\nonumber\\
    &\psi^{\dagger}_{a}(\textbf{x}) = \sum_{s=1}^4 \int \frac{d^3\textbf{p}}{(2\pi)^{3/2}}\  \frac{1}{\sqrt{\omega_{\textbf{p}}}} b^{\dagger}(\textbf{p},s)u^{\dagger}_a(\textbf{p},s)e^{-i\textbf{p}\cdot\textbf{x}},
\end{align}

where the spinors $u(\textbf{p},s)$, $s = 1,...4$ satisfy:

\begin{align}
    &(\slashed{p} - m)u(\textbf{p},s) = 0 \Rightarrow\nonumber\\ &(\textbf{p}\cdot \vec{\gamma} + m)u(\textbf{p},s) = p_0\gamma^0 u(\textbf{p},s) = \omega_{\textbf{p}}\gamma^0 u(\textbf{p},s),\nonumber\\
    &u^{\dagger}(\textbf{p},s)u(\textbf{p},r) = \omega_{\textbf{p}}\delta_{sr},\nonumber\\
    &\overline{u}(\textbf{p},s)u(\textbf{p},r) = \sigma_s \omega_{\textbf{p}}\delta_{sr}.
\end{align}

Here, $\sigma_s = 1$ if $s=1,2$ and $\sigma_s = -1$ if $s = 3,4$. With this, the Dirac Hamiltonian (\ref{Dirac Hamiltonian}) is written as:

\begin{equation}\label{Hamiltonian b}
    H = \hbar\sum_{s = 1}^4 \int d^3\textbf{p}\ \sigma_s \omega_{\textbf{p}} b^{\dagger}(\textbf{p},s)\star b(\textbf{p},s).
\end{equation}

Also, the star product can be expressed in terms of $b,b^{\dagger}$:

\begin{align}
    &b(\textbf{p},s)\star = b(\textbf{p},s) + \frac{1}{2}\partialfunc{}{b^{\dagger}(\textbf{p},s)}, \nonumber\\ &b^{\dagger}(\textbf{p},s)\star = b^{\dagger}(\textbf{p},s) + \frac{1}{2}\partialfunc{}{b(\textbf{p},s)},
\end{align}

which means that the anticommutation relations are now:

\begin{align}
    &\{\{ b(\textbf{p}, s), b(\textbf{p}',r)\}\}_{anti} = \{\{ b^{\dagger}(\textbf{p}, s), b^{\dagger}(\textbf{p}',r)\}\}_{anti} = 0, \nonumber\\ &\{\{ b(\textbf{p}, s), b^{\dagger}(\textbf{p}',r)\}\}_{anti} = \delta_{rs}\delta^{(3)}(\textbf{p}-\textbf{p}').
\end{align}

It can be easily seen that the functionals $b(\textbf{p},s=1,2)$ correspond to particles of positive energy, while \\$b(\textbf{p},s=3,4)$ correspond to particles of negative energy. The label $s$ is related to the spin of the particle.

With this information, we are ready to calculate the fundamental state of the free fermionic field. We start noting that this state is not empty, instead, all the negative energy states are occupied; this is as the Dirac sea. Therefore, one must construct a Wigner functional in which all states with positive energy are vacuum and all states with negative energy are filled. These conditions translate into the following relations:

\begin{equation}\label{fermionic Wigner conditions}
\begin{gathered}
    b(\textbf{p},s=1,2)\star W_0[b,b^{\dagger}] = W_0[b,b^{\dagger}]\star b^{\dagger}(\textbf{p},s=1,2) = 0,\\
    b^{\dagger}(\textbf{p},s=3,4)\star W_0[b,b^{\dagger}] = W_0[b,b^{\dagger}]\star b(\textbf{p},s=3,4) = 0.
\end{gathered}
\end{equation}

Indeed, since the states with positive energy are empty, annihilating any of them cancels the functional. In the same way, since the states with negative energy are filled, creating any of them also cancels the vacuum functional, because there cannot be two fermions in the same state. 

To simplify the calculation, let us assume that only one possible state exists. If it is empty, then the first of the conditions (\ref{fermionic Wigner conditions}) must be satisfied, that is:

\begin{align}
    &b\, W_{e}[b,b^{\dagger}] = -\frac{1}{2}\partialfunc{W_{e}}{b^{\dagger}}, \nonumber\\ &W_{e}[b,b^{\dagger}]\, b^{\dagger} = -\frac{1}{2}W_{e}[b,b^{\dagger}]\frac{\overleftarrow{\delta}}{\delta b}\,.
\end{align}

The solution to these equations is given by:

\begin{equation}
    W_{e}[b,b^{\dagger}] = 1 + 2bb^{\dagger} = \exp(2bb^{\dagger}),
\end{equation}

where, of course, $b$ and $b^{\dagger}$ are Grassmann variables. On the other hand, if the state is occupied, then:

\begin{align}
    &b^{\dagger}\, W_{o}[b,b^{\dagger}] = -\frac{1}{2}\partialfunc{W_{o}}{b},\nonumber \\ &W_{o}[b,b^{\dagger}]\, b = -\frac{1}{2}W_{o}[b,b^{\dagger}]\frac{\overleftarrow{\delta}}{\delta b^{\dagger}}.
\end{align}

In this case, the Wigner functional is:

\begin{equation}
    W_o[b,b^{\dagger}] = 1 + 2b^{\dagger}b = \exp(2b^{\dagger}b).
\end{equation}

As can be seen, the difference between an occupied and an empty state is simply the order in which $b$ and $b^{\dagger}$ appear in the Wigner functional. Extrapolating from this result, it follows that the Wigner functional which represents the vacuum state for the free fermionic field is:

\begin{align}
    &W_0[\psi,\psi^{\dagger}] = \mathcal{N}\exp\left(2\int d^3\textbf{p} \ \sum_{s=1,2}b(\textbf{p},s)b^{\dagger}(\textbf{p},s) + \right. \nonumber \\ &\left. \sum_{r=3,4} b^{\dagger}(\textbf{p},r)b(\textbf{p},r)\right).
\end{align}

The energy of this state can be calculated using the expression for the Hamiltonian (\ref{Hamiltonian b}):

\begin{align}
    H\star W_0 = E_0\,W_0 \Rightarrow E_0 =  -2\int d^3\textbf{p}\ \omega_{\textbf{p}}\delta^{(3)}(\textbf{0}). 
\end{align}

This is consistent with the idea that the Dirac sea is occupied by all possible particles with negative energy.

Next, in order to create states with positive energy particles (which are the ones that can be observed experimentally), we simply apply the creation/annihilation functionals. For example, the state of a fermion with momentum $\textbf{q}$ and spin $+1/2$ is represented by:

\begin{align}
    &b^{\dagger}(\textbf{q},1)\star W_0[b,b^{\dagger}] \star b(\textbf{q},1)  =\mathcal{N}\exp\left( 2b^{\dagger}(\textbf{q},1)b(\textbf{q},1)  \right) \nonumber\\&\exp\left(2\int_{(\textbf{p},s) \neq (\textbf{q},1)} d^3\textbf{p} \ \sum_{s=1,2}b(\textbf{p},s)b^{\dagger}(\textbf{p},s) + \right. \nonumber\\ &\left. \sum_{r=3,4} b^{\dagger}(\textbf{p},r)b(\textbf{p},r)\right).
\end{align}

That is, one must change the order of the product $b^{\dagger}b$ for that particular momentum and spin. On the contrary, to create an antifermion with momentum $-\textbf{q}$ and spin $-1/2$, it is necessary to destroy one of the particle states with negative energy (namely, the particle state with momentum $\textbf{q}$ and spin $+1/2$), since that gap is interpreted as the presence of a positive energy antiparticle. Thus, the Wigner functional for this type of state is given by:

\begin{align}
    &b(\textbf{q},3)\star W_0[b,b^{\dagger}]\star b^{\dagger}(\textbf{q},3)  =\mathcal{N}\exp\left( 2b(\textbf{q},3)b^{\dagger}(\textbf{q},3)  \right) \nonumber\\&\exp\left(2\int_{(\textbf{p},s) \neq (\textbf{q},3)} d^3\textbf{p} \ \sum_{s=1,2}b(\textbf{p},s)b^{\dagger}(\textbf{p},s) + \right.\nonumber \\ &\left. \sum_{r=3,4} b^{\dagger}(\textbf{p},r)b(\textbf{p},r)\right).
\end{align}

The next step is to verify that these states represent particles with positive energy. Applying the Hamiltonian, we obtain:

\begin{align}
    &H\star  b^{\dagger}(\textbf{q},1)\star W_0 \star b(\textbf{q},1) =\nonumber\\ &(E_0 + \hbar \omega_{\textbf{q}}) b^{\dagger}(\textbf{q},1)\star W_0 \star b(\textbf{q},1),
\end{align}
\begin{align}
     &H\star  b(\textbf{q},3)\star W_0 \star b^{\dagger}(\textbf{q},3) =\nonumber\\ &(E_0 + \hbar \omega_{\textbf{q}}) b^{\dagger}(\textbf{q},3)\star W_0 \star b(\textbf{q},3),
\end{align}

where we have used anti-commutation rules for the creation and annihilation functionals. As expected, the states that have been characterized as fermions and antifermions contribute with a positive energy to the field. To show that they correspond to particle and antiparticle, respectively, one can calculate the charge functional, which in fermionic field theory is given by:

\begin{equation}
\begin{gathered}
    Q = \int d^3\textbf{x}\ \psi^{\dagger}(\textbf{x})\star \psi(\textbf{x}) = \sum_{s = 1}^4 \int d^3\textbf{p}\ b^{\dagger}(\textbf{p},s)\star b(\textbf{p},s).
\end{gathered}
\end{equation}

Using this expression, a direct calculation shows that the fermion and antifermion states are eigenstates of $Q$, with eigenvalues $Q_0 + 1$ and $Q_0 - 1$, respectively. Here, $Q_0 = 2\int d^3\textbf{p}\ \delta^{(3)}(\textbf{0})$ is the charge of the fundamental state. It is thus evident that the particle states have charge $+1$, while the antiparticle states have charge $-1$. We can also confirm that these states are associated with the corresponding momentum and spin (the computations are completely analogous to those performed for this and the previous fields). In this way, the free fermionic field has been completely characterized in phase space.
Note that the interpretation of the Dirac sea used for the calculation of the Wigner functional is just that, an interpretation. However, functionals characterized in this way perfectly describe well-defined states of particles and antiparticles of positive energy. If we wanted to eliminate the energy and charge of the "vacuum", we would simply have to renormalize using, for example, normal ordering in the functionals.

\section{Interaction theories}

Once the Wigner formalism for free fields has been developed, the next step would be to extend it to theories where different fields interact with each other, such as quantum electrodynamics or self-interacting scalar $\phi^4$ theory. The procedure would be analogous to the one followed in the Schrödinger picture: the Hamiltonian is expressed as the sum of a free contribution and a sufficiently small interaction term $H = H^{0} + \lambda H^{1}$, with $\lambda \ll 1$, and the self-states and self-energies are expanded in powers of $\lambda$,

\begin{align}
    &W_n[\phi,\pi] = W_n^0[\phi,\pi] + \lambda W_n^1[\phi,\pi] + \lambda^2 W_n^2[\phi,\pi] +..., \nonumber\\ \ &E_n = E_0 + \lambda E_1 + \lambda^2 E_2+...
\end{align}

Imposing the eigenvalue equation $H\star W_n = W_n\star H = E_nW_n$ and identifying the terms with the same power of $\lambda$, we arrive at \cite{Perturbation}:

\begin{align}
    &E_0 = \int \mathcal{D}\phi \mathcal{D}\pi\ H^0[\phi,\pi]W_n^0[\phi,\pi], \\
    &E_1 = \int \mathcal{D}\phi \mathcal{D}\pi\ H^1[\phi,\pi]W_n^0[\phi,\pi],\\
    &W_n^1[\phi,\pi] = \nonumber\\ &\sum_{m\neq n} \frac{1}{E_n^0-E_m^0} W_n^0[\phi,\pi] \int \mathcal{D}\phi \mathcal{D}\pi\ H^1[\phi,\pi]W_n^0[\phi,\pi] +  \nonumber\\ &\sum_{m\neq n} W_n^0[\phi,\pi] \int \mathcal{D}\phi \mathcal{D}\pi\ H^1[\phi,\pi]W_n^0[\phi,\pi]... 
\end{align}

Thus, solving by iteration, the Wigner functional is obtained up to the desired order. As straightforward as this procedure seems, we have to take into account that all the integrals to be performed are functional integrals, most of which do not present immediate solutions. Moreover, since each term of the series $\lambda^m W_n^m[\phi,\pi]$ is associated with a diagram with $m$ vertices, the number of orders that need to be calculated for most of the processes becomes unreasonably large, especially since each order is determined from the previous ones. One can appreciate that, even in the Schrödinger formalism, the calculations become unnecessarily complicated compared to other procedures such as the path integral \cite{Hatfield}. Therefore, it is not difficult to see that the Wigner formalism, where the inclusion of the conjugate momenta $\pi(\textbf{x})$ effectively doubles the length and difficulty of the various computations, will not be able to compete with the more efficient formalisms either when determining transition amplitudes and corrections to the energy.

\section{Conclusions}

In this work, we developed the quantum theory of free fields in the Wigner formalism. In order to do this, we used the so-called functional representation of QFT, according to which the physical states of a system are described by functionals of the fields. In this case, the distribution function is defined in infinite-dimensional phase space $(\phi(\textbf{x}),\pi(\textbf{x}))$, and the various observables are functionals of these fields, $A[\phi,\pi]$. As in the finite-dimensional case, the Wigner transform (\ref{Wigner transform QFT}) and the star product (\ref{star product QFT}) are defined, and the stationary states are obtained from the eigenvalue equation, which is now a functional differential equation.

In this way, we determined the fundamental and first excited states of the real and complex scalar field, the fermionic field and the massless electromagnetic field theories. We found them to be consistent with the particle states obtained in the Heisenberg picture (which is, together with the Feynman path integral, the standard formulation of quantum field theory). The calculations of these states, although feasible, were rather tedious due to the functional nature of the theory, especially those of the fermionic field, where it was necessary to use Grassmann variables for the field functionals to guarantee the anticommutation relations. It should be noted that these free fields are the simplest systems to analyze, since they are described by infinite-dimensional harmonic oscillators. Once interaction terms are introduced into the theory, computations become significantly more complicated in comparison to usual procedures.

In summary, functional phase space representations of QFT provide an alternative and perhaps more intuitive approach to different physical systems. However, for a systematic perturbative treatment of interactions, the increased difficulty in the calculations make them less efficient for obtaining analytical and numerical results. Particular applications for other systems will be analysed in future work.

\begin{acknowledgements}
We would like to thank Cosmas K. Zachos for fruitful conversations about the functional representations of Quantum Field Theory in Phase Space.
This work was partially supported by Grant 512590 (MEFP, Spain) and the MINECO (Spain) project PID2019-107394GB-I00 (AEI/FEDER, UE).
\end{acknowledgements}

\end{document}